\documentclass{ws-procs9x6}

\newcommand{\be}{\begin{equation}}
\newcommand{\ee}{\end{equation}}
\newcommand{\bea}{\begin{eqnarray}}
\newcommand{\eea}{\end{eqnarray}}
\newcommand{\hf}{\frac{1}{2}}

\def\Journal#1#2#3#4{{#1}{\bf #2},#3(#4)}

\def\eq#1{(\ref{#1})}

\def\mr#1{\mathrm{#1}}

\def\la{\langle}
\def\ra{\rangle}

\begin{document}
\title{Spinodal Instability and Confinement
\footnote{\uppercase{T}alk presented at the Workshop \uppercase{GRIBOV-75},
\uppercase{B}udapest, \uppercase{M}ay 2005.}}
\author{Janos Polonyi}
\address{Institute for Theoretical Physics, Louis Pasteur University, Strasbourg, France}
\maketitle

\abstracts{It is pointed out that inhomogeneous condensates or
spinodal instabilities suppress the propagation of elementary excitations
due to the absorptive zero mode dynamics. This mechanism is shown to be present
in the scalar $\phi^4$ model and in Quantum Gravity. It is conjectured
that the plane waves states of color charges have vanishing
scattering amplitude owing to the color condensate in the vacuum.}

\section{Introduction}
The mechanism of quark confinement continues to be one of the stumbling
block in particle physics and one tends to shelve it by the help
of more or less ad hoc models motivated by QCD. Instead of
a detailed construction of "The Mechanism" which obviously involves some
highly technical and non-perturbative issues of QCD we suggest in this
talk a rather well known phenomenon, spinodal instability, as a natural
ingredient of confinement.

There are actually two mechanisms for quark confinement. One is
established by means of studying Wilson-loops in
quenched lattice QCD without dynamical quarks and is
operating with a linearly rising potential between static fundamental
color charges\cite{wilson}. One may call this hard mechanism because it is
based on large energy scales to suppress localized quarks.
Another mechanism, proposed by V. Gribov\cite{gribov}, is conjectured
in QCD with dynamical, light quarks and is reminiscent of the
supercritical vacuum in QED. The separation of a quark form a hadron
triggers an increase in the running coupling constant which in turn
generates screening by the anti-quark of a $q-\bar q$ vacuum-polarization.
We may call this soft mechanism because the screening is achieved
by vacuum fluctuations of the energy of a light meson.
The soft mechanism is the real one, observed in
hadrons and is based on the non-perturbative phenomena in the
Dirac-see caused by the increase of the running coupling constant
at large distances, the key phenomenon of the hard mechanism.
An inherent difficulty of both mechanisms is the explanation of gluon
confinement. We argue below that the colored condensate of
the QCD vacuum should drive a spinodal instability which
automatically confines color charges, both gluons and quarks.

We shall discuss this mechanism in the framework of two theories.
One is the the $\phi^4$ scalar model in the symmetry broken phase.
We argue that when a homogeneous external source is coupled to the
field $\phi(x)$ and its value is chosen in such a manner that
$|\la\phi(x)\ra|$ is decreased then the elementary excitations
become confined. The second example is Quantum Gravity where
confinement of gravitons is expected.
QCD is touched upon briefly at the end only.

\begin{figure}[ht]
\centerline{\epsfxsize=5cm\epsfbox{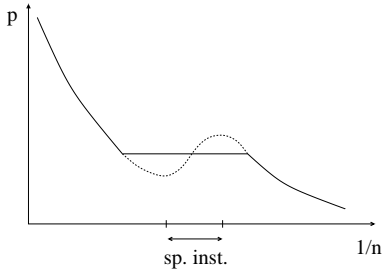}}
\caption{A qualitative sketch of the pressure $p$ as the function of the
density $n$ for the van der Waals equation of state. The density 
interpolates between the values corresponding to the two phases 
while the pressure remains unchanged on the horizontal line, 
the Maxwell-cut.\label{vdw}}
\end{figure}

The basic idea is very simple. Let us start with the spinodal instability 
shown in Fig. \ref{vdw} which is usually observed in first order phase 
transitions.The spinodal unstable region is characterized by the apparent 
violation of 
the sum rule which assures that the pressure is a non-decreasing function of 
the density. The Maxwell-cut restores stability and renders the pressure
independent of the density in the mixed phase where domains
made up by the two stable phases are formed. The sound velocity,
\be
c=\sqrt{-\frac{1}{n^2}\frac{\partial p}{\partial1/n}},
\ee
is clearly vanishing in the mixed phase. The vanishing is due to the
domain walls which can be displaced without energy absorb the density
waves. The density fluctuations are therefore non-propagating 
and the sound waves become "confined".

\section{Scalar model}
\begin{figure}[ht]
\centerline{\epsfxsize=5cm\epsfbox{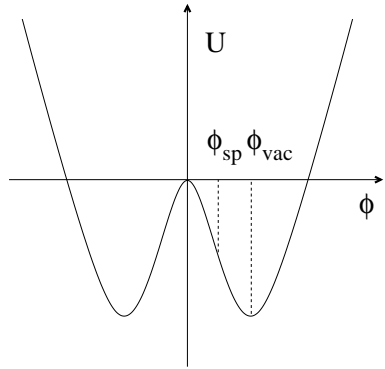}}
\caption{The double-well potential of the scalar model.\label{pot}}
\end{figure}

One can gain more insight into the dynamics of the instabilities of 
Fig. \ref{vdw} by considering a scalar model where a homogeneous external 
source, $j$, coupled linearly to the field $\phi(x)$ is introduced in 
order to dial a desired value of the condensate. The partition function of the
Euclidean model is 
\be
Z=\int D[\phi]e^{-\int dx[\hf(\partial\phi(x))^2+U(\phi(x))]+j\int dx\phi(x)}
\ee
where the potential is an even function, $U(-\phi)=U(\phi)$, 
and is double-well shaped, as shown in Fig. \ref{pot}. We shall determine
the vacuum which is supposed to be homogeneous for any 
value of the source $j$ in the tree-level approximation.
The vacuum of the theory with $j=0$ is dominated by a single homogeneous
configuration  $\la\phi(x)\ra=\phi_\mr{vac}>0$
satisfying the conditions $U'(\phi_\mr{vac})=0$ and 
$U''(\phi_\mr{vac})>0$.

\begin{figure}[ht]
\centerline{\epsfxsize=5cm\epsfbox{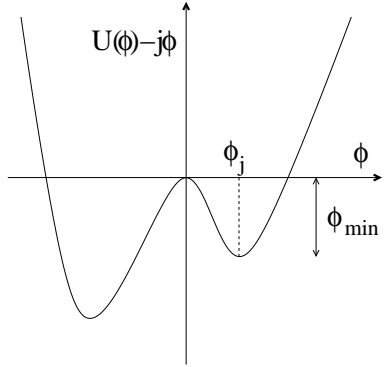}}
\caption{The vacuum is dominated by a single homogeneous saddle point in the
nucleation phase, given by Eq. \eq{nuclph}.\label{nucl}}
\end{figure}

The gradual decrease of the external source towards more negative
values decreases the saddle-point $\phi(x)=\phi_j>0$ which is defined by 
the equation $U'(\phi_j)=j$, as shown in Fig. \ref{nucl}.
The homogeneous saddle-point vacuum remains stable, $U''(\phi_j)>0$, as long as 
\be\label{nuclph}
\phi_\mr{sp}\le\phi_j\le\phi_\mr{vac},
\ee
where the lower bound
is defined by the equation $U''(\phi_\mr{sp})=0$. Notice that the
vacuum constructed in such a manner is stable against infinitesimal
fluctuations but decays into the the true vacuum, given by
the absolute minimum of $U(\phi)+j\phi$ when fluctuations with
sufficiently large amplitude, $\Delta\phi>\phi_\mr{min}$ are formed. 
This regime of the model
is called nucleation phase because the large amplitude fluctuations
which destabilize the false vacuum are energetically stable above
a certain size. All modes are massive in this phase.

\begin{figure}[ht]
\centerline{\epsfxsize=5cm\epsfbox{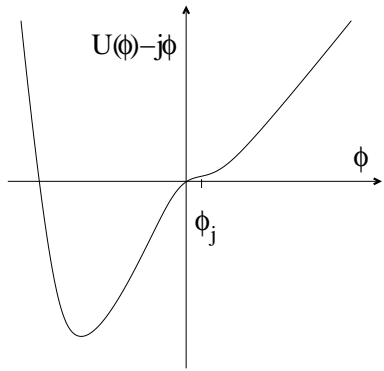}}
\caption{The potential energy in the spinodal phase.\label{spinod}}
\end{figure}

As the external source is pushed further towards the negative direction
the saddle-point reaches the lower bound in Eq. \eq{nuclph} and 
we enter into the spinodal instability region, cf. Fig. \ref{spinod},
characterized by instabilities against infinitesimal fluctuations.
The saddle-points are inhomogeneous and support domains where $\phi(x)$
assumes opposite sign. The breakdown of the external, space-time symmetries
induces zero modes, the location and the direction of the domain walls. We shall argue in 
the next Section that these soft modes make the "sound waves" 
non-propagating, ie. remove the plane waves from the asymptotical 
scattering states of the theory.

\section{Instability induced renormalization}
The inhomogeneous saddle-points appearing in the spinodal phase require
more powerful method than the mean-field approximation. It is a time 
honored strategy to deal with modes one-by-one in a sequential manner,
as in the renormalization group method\cite{rg}, instead of facing all of them
in the same time. Such a scheme proves to be useful for the
saddle-points, too\cite{iir}. 

\begin{figure}[ht]
\centerline{\epsfxsize=5cm\epsfbox{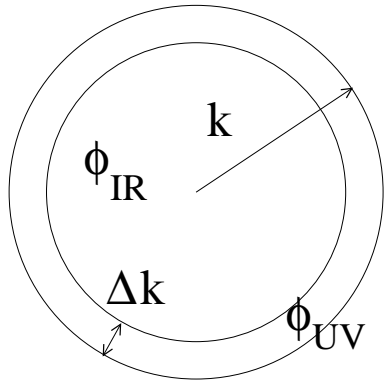}}
\caption{The split of the field variable into an infrared and ultraviolet
component is defined in the Fourier space.\label{bril}}
\end{figure}

Let us consider an infinitesimal decrease of the UV cutoff, 
$k\to k-\Delta k$, in momentum space which induces 
the change $S_k[\phi]\to S_{k-\Delta k}[\phi]$ in the bare action
of the scalar model,
\be\label{eveq}
e^{-\frac{1}{\hbar}S_{k-\Delta k}[\phi_\mr{IR}]}
=\int D[\phi_\mr{UV}]e^{-\frac{1}{\hbar}S_k[\phi_\mr{IR}+\phi_\mr{UV}]},
\ee
where the split $\phi(x)=\phi_\mr{IR}(x)+\phi_\mr{UV}(x)$
of the field variable is defined by requiring that the Fourier
transform of $\phi_\mr{UV}(x)$ is non-vanishing within the shell
$k-\Delta k<p<k$ only, as shown in Fig. \ref{bril}. The tree-level
contribution to the evolution equation \eq{eveq} is
\be
S_{k-\Delta k}[\phi_\mr{IR}]=S_k[\phi_\mr{IR}+\phi_\mr{UV}[\phi_\mr{IR}]].
\ee
A non-trivial saddle-point, $\phi_\mr{UV}[\phi_\mr{IR}]\not=0$, induces an
evolution in the bare action, $S_{k-\Delta k}[\phi_\mr{IR}]\not=S_k[\phi_\mr{IR}]$.

Two remarks are in order now about the saddle-points. The first is that 
the non-vanishing saddle-point is the hallmark of spontaneous symmetry 
breaking, the appearance of a condensate in the theory. In fact,
the inverse propagator, $\delta^2S_k[\phi]/\delta\phi\delta\phi$, 
approaches its renormalized form as $k\to0$. Spontaneous
symmetry breaking is driven by the negative values of the renormalized 
propagator for small momenta computed in the symmetrical vacuum, $\phi(x)=0$.
Hence $\phi(x)=0$ ceases to be the absolute minimum of the bare action 
at sufficiently small values of the cutoff $k$ in the symmetry broken
phase and a non-trivial saddle point is found for Eq. \eq{eveq}.
The other remark is about the circumstance that these non-trivial 
saddle-points appear in a scalar model in
any dimension even if the whole partition function supports no
solitons or instantons. The saddle-points mentioned again and again in
this Section correspond to a constrained functional integral where
the blocked variables are held fixed. It is this constrain which induces
the block variable-dependent saddle-points and the partition function
as a complicated integral does not reveal their existence.

Let us find the tree-level solution of the evolution equation in the
local potential approximation where the bare action is assumed to
be of the form
\be
S_k[\phi]=\int dx\left[\hf(\partial\phi(x))^2+U_k(\phi(x))\right].
\ee
This ansatz allows us to seek the solution only for $\phi_\mr{IR}(x)=\Phi$ because
\be\label{min}
U_{k-\Delta k}(\Phi)=\min_{\phi_\mr{UV}}\int dx\left[
\hf(\partial\phi_\mr{UV})^2+U_k(\Phi+\phi_\mr{UV}(x))\right].
\ee
As a further simplification, we search for the minimum among the 
plane wave configurations only,
\be\label{saddlep}
\phi_\mr{UV}(x)=\rho_k\cos(kn_k\cdot x+\theta_k).
\ee
Note that the unit vector $n_k$ and the phase $\theta_k$ are zero
modes arising from the breakdown of the rotational and translational
symmetry. Eq. \eq{min} now reads as
\be\label{plev}
U_{k-\Delta k}(\Phi)=\min_{\rho_k}\left(\frac14k^2\rho_k^2
+\frac{1}{\pi}\int_0^\pi dyU_k(\Phi+\rho_k\cos y)\right).
\ee

\begin{figure}[ht]
\centerline{\epsfxsize=5cm\epsfbox{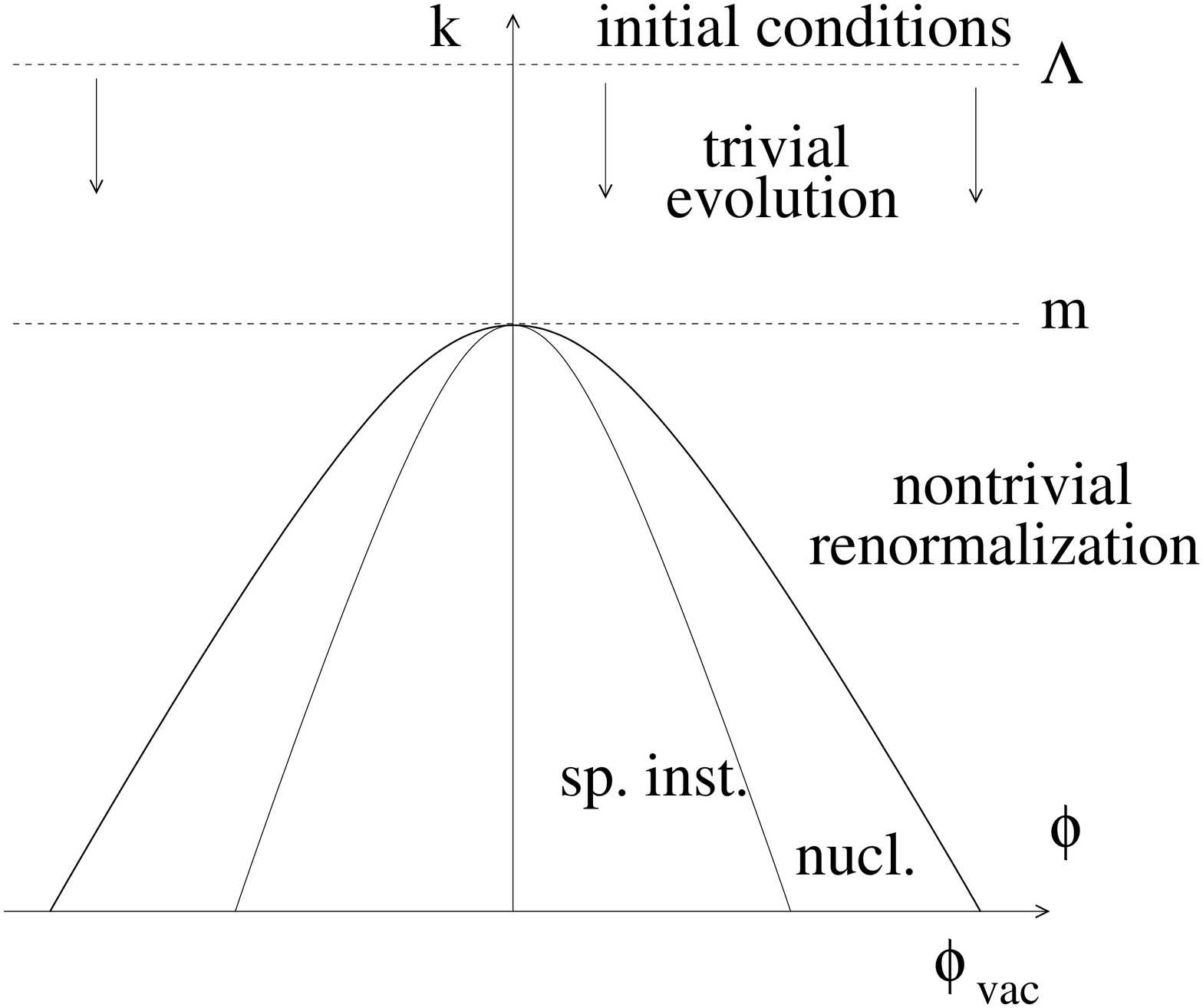}}
\caption{Tree-level evolution of the potential $U(\phi)$.
The initial condition is set at $k=\Lambda\gg\sqrt{-m^2}$ but the 
tree-level evolution starts only for $k^2=-m^2$. The saddle-point
is non-vanishing within the area bounded by the outer curve. The
naive reasoning would suggest the possibility of separating
nucleation and spinodal phases in the instable regime. This proved
to be wrong by the numerical solution, the spinodal region spreads
over the whole unstable domain.\label{trg}}
\end{figure}

In order to obtain the renormalization group flow we impose the 
initial condition
\be
U_k(\phi)=-\frac{m^2}{2}\phi^2+\frac{g}{4!}\phi^4,
\ee
with $m^2<0$ at $k=\Lambda\gg\sqrt{-m^2}$ for Eq. \eq{plev}. There is no 
evolution for $k\approx\Lambda$ because the kinetic energy dominates
the inverse propagator
\be
G(p)=\frac{1}{p^2-m^2+\frac{g}{2}\Phi^2}
\ee
for $p^2\gg-m^2$, cf. Fig. \ref{trg}. The inhomogeneous saddle-point 
becomes non-trivial and the potential evolves for $k^2<-m^2$.

The following observations can be made after inspecting 
the numerical solution of Eq. \eq{plev}:
\begin{enumerate}
\item Eq. \eq{plev} goes over a differential equation in the limit
$\Delta k\to0$ and $S_k[\phi]$ is continuous\cite{regrg} in $k$.

\item The instability is always followed by
the appearance of inhomogeneous saddle-points and their zero modes.
The zero modes are soft and can be excited by arbitrarely
small energies, indicating spinodal phase separation in the whole
instability region. The nucleation phase expected naively in
between the stable and the spinodal region disappears.

\item The bare action $S_k[\phi]$ is always flat for modes with $p=k$,
in the variable $\phi_\mr{UV}$, due to the cancellation between the
kinetic energy and the potential energy which is found to be
\be\label{iepot}
U_k(\Phi)=-\hf k^2\Phi^2
\ee
within the instability region. The resulting flat action,
the lack of restoring force to any equilibrium position,
becomes the usual Maxwell-cut as $k\to0$.

\item The vacuum, $\phi_\mr{vac}$, is at the edge of the unstable region.

\item These tree-level results, Eq. \eq{iepot} in particularly, hold
independently of the choice of the potential at the cut-off.
\end{enumerate}

The unexpected feature is the appearance of zero modes, $n_k$ and $\theta_k$
in Eq. \eq{saddlep}, within the 
unstable region in the one-component model with the discrete internal 
symmetry $\phi(x)\to-\phi(x)$. These are Goldstone modes arising from the
breakdown of translational and rotational symmetry by the inhomogeneous 
saddle-points. The plane-wave is a poor-man's approach to the domain wall 
structure and its phase and direction control the location
and the direction of the walls. The integration over the zero modes 
appearing independently at each value of $k$ restores the external
symmetries and renders the vacuum homogeneous.

It is point 3 which is relevant from the point of view of confinement. 
What we have obtained here is a layer-by-layer scan of the
bare action for this theory. In fact, the dynamics of a mode with wave
vector $p$ is not contained explicitly or appears in a rather approximative
manner only in $S_k[\phi]$ for $|p|>k$ or $|p|<k$. This dynamics is best
described by the bare action with $k=|p|$.
Let us suppose that the external homogeneous source stabilizes the vacuum at
$|\la\phi(x)\ra|<\phi_\mr{vac}$ and the model is in the spinodal
phase. The cancellation between the
kinetic and the potential energy should therefor be kept
as a feature characterizing all unstable modes and no plane waves
exist with small wave numbers. One may say in the more formal language
of the reduction formulae that there are no mass-shell singularities in
the Green functions for these modes and the scattering amplitude is
vanishing, ie. the low energy elementary quanta of the theory are
non-propagating, confined. 

What happens with the high energy modes which belong to the stable
regime of Fig. \eq{trg}? These modes are supposed to propagate
on the background field of the inhomogeneous condensate and the
integration over the zero modes appears as a quenched average.
One expects that this is a strongly disordered system where
the Euclidean analogy of the Anderson localization prevents
these states from being extended. In the lack of a detailed
study it is sufficient to say that the restored external symmetries
impose the cancellation of the scattering amplitude at any energy
once it is found vanishing at low energy. In other words, all particle
modes are confined as long as there is a finite low energy range
with inhomogeneous condensate.

\section{Savvidy vacuum in Quantum Gravity}
There is another model where confinement of the elementary excitations
are expected in the semiclassical approximation, namely gravity\cite{salsat}.
Let us consider a point-like body surrounded by its horizon and suppose that
the position of the body undergoes small amplitude oscillations. The
acceleration sends gravitational waves into the space-time within the horizon.
Though the non-propagating features of the solution of the Einstein equations
may indicate the motion of the body outside of the horizon, the
propagating gravitons are supposed to remain confined by the horizon.

It is not difficult to recognize the trace of spinodal instability
in quantum gravity. Let us consider the partition function\cite{pore}
\be
Z=\int D[g]e^{-S_E[g]},
\ee
for Euclidean signature metric tensor, where the Einstein action is given by
\be\label{eact}
S_E[g]=-\kappa^2_B\int dx\sqrt{g}R.
\ee
This theory poses serious problems both in the UV and the IR domains.

The difficulty in the UV region is that the theory is non-renormalizable.
We shall keep the cutoff $\Lambda$ at a large but finite value and consider
\eq{eact} as an effective action valid for distances
$\ell>\ell_\mr{min}\approx1/\Lambda$ only. The regularization is carried
out by restricting the functional integral for geometries into the
vicinity of de Sitter spaces, characterized by their curvature $R$,
and by suppressing modes whose eigenvalue with respect to the 
operator $-D^2$ is superior to $\Lambda^2$.

The IR problems come from the fact
the the action \eq{eact} is unbounded form below in the conform modes,
$\Omega$, defined as $\tilde g_{\mu\nu}=g_{\mu\nu}\Omega^2$
because
\be
S_E[\tilde g]=-\kappa^2\int dx\sqrt{g}
\left(\Omega^2 R+6\Omega_{;\mu}\Omega_{;\nu}g^{\mu\nu}\right).
\ee
Instead of modifying the Wick-rotation\cite{hawk}, the quantization rules
\cite{quantst} or the functional integral measure\cite{fume} which lead 
to further problems with general covariance it is more illuminating
to recognize the similarity to the instability of the 
Yang-Mills vacuum\cite{savv}. 
In fact, the interactions among the elementary excitations 
are attractive in both cases and populate modes macroscopically.
These modes are inhomogeneous in order not to break gauge and external
symmetries. The apparent difference, the Yang-Mills action is
bounded from below but the gravitational one is not, may not
be important because the stabilization of the vacuum can
be achieved by quantum fluctuations. An obvious example
is the Hydrogen atom where the kinetic energy created by
localization balances the attractive, unbounded Coulomb 
potential.

\begin{figure}[ht]
\centerline{\epsfxsize=5cm\epsfbox{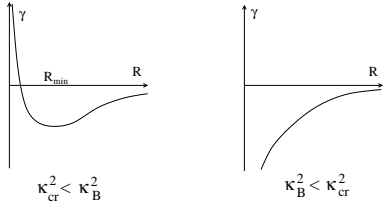}}
\caption{The effective potential as the function of the de Sitter
curvature in the low and the high cutoff phases.\label{greff}}
\end{figure}

The one-loop contributions of the stable modes to the effective potential
$\gamma_\mr{eff}(R)=-\ln Z(R)$ where $Z(R)$ is the perturbative
partition function on the de Sitter space of curvature $R$ give
\be\label{effp}
\gamma(R)=\begin{cases}-\frac{v\kappa^2_B}{R}
+c_1\Lambda^4\ln\frac{c_2\kappa^2_B}{\Lambda^2}
\left(\frac1{R^2}-\frac1{c^2_3\Lambda^4}\right)&R\le c_3\Lambda^2\cr
-\frac{v\kappa^2_B}{R}&R>c_3\Lambda^2\end{cases},
\ee
with $c_1\approx7.201$, $c_2\approx2.989$ and $c_3\approx0.665$,
cf. Figs. \ref{greff}. This result indicates a quantum 
phase transition at $\kappa_{cr}^2=\frac{\Lambda^2}{c_2}$ and the curvature
\be\label{vacr}
R_{min}=\begin{cases}\frac{2c_1\Lambda^4}{v\kappa^2_B}\ln\frac{c_2\kappa^2_B}{\Lambda^2}
&\kappa_{cr}^2\ll\kappa^2_B\cr0&\kappa^2_B\ll\kappa_{cr}^2\end{cases}
\ee
at the minimum of the effective potential. What is left to see whether
the conformal modes are stabilized at these vacua. This phase diagram
agrees qualitatively with the one found in lattice simulations\cite{latt}
suggesting a radiative corrections induced stabilization of the vacua of 
Eqs. \eq{vacr}. Notice that whatever stabilization mechanism prevails 
the vacuum contains inhomogeneous condensate in both phases, made of the
inhomogeneous conformal modes, and one expects that the appearing zero 
modes make the plane waves, gravitons, non-propagating.

\section{Conclusions}
It has been argued that inhomogeneous condensate, in particular
spinodal instability, leads to the confinement of elementary 
excitations by the absorptive dynamics of the zero modes. 
Two models have been mentioned, a scalar $\phi^4$ model and 
Euclidean quantum gravity. 
The mechanism is established on a qualitative level only,
the more satisfactory construction obviously requires 
a systematical and through study of the tree-level 
contributions to the renormalization group equations.

\begin{figure}[ht]
\centerline{\epsfxsize=5cm\epsfbox{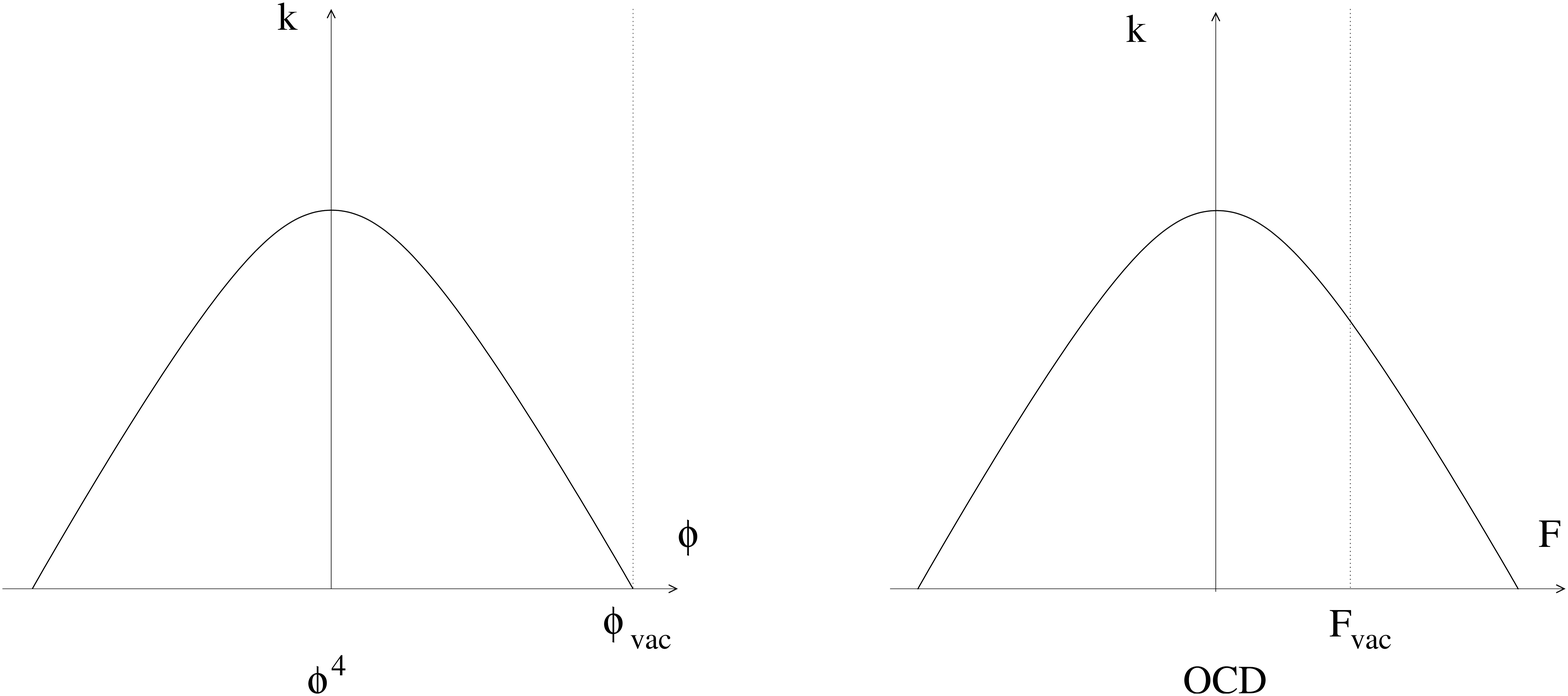}}
\caption{The tree-level evolution in the $\phi^4$ model (left) and
in QCD (right). The vertical axes show the cutoff, the horizontal ones
correspond to the expectation value of the scalar field and
some components of the field strength tensor. The
inhomogeneous saddle-points appear within the regions
bounded by the solid line.\label{qcd}}
\end{figure}

Is this mechanism effective in the QCD vacuum? Let us compare
the known tree-level renormalization flow of the scalar model with the 
expected flow in QCD, as shown in Figs. \ref{qcd}. The vacuum of the $\phi^4$
model, shown in the first figure, is at the edge of the spinodal
phase. The effects of the eventual zero modes within the unstable 
region are suppressed and no confinement occurs in the absence of the
external source. The color condensate 
disappears at short distances or in the presence of strong external 
chromomagnetic field in QCD. The region with condensate is shown 
qualitatively in the second figure. What is important is
that the true vacuum is within the unstable region. Hence
one expects that the zero mode dynamics of the inhomogeneous
condensate renders the effective action flat and removes colored
plane wave states from the asymptotical sector of QCD.

\end{document}